\documentclass[12pt,pre,aps,superscriptaddress]{revtex4}
\usepackage{graphicx}
\usepackage{bm}
\usepackage{amssymb}
\newcommand{\beq}[2]{\begin{equation}#1\label{#2}\end{equation}}
\newcommand{\ceq}[1]{(\ref{#1})}

\newfont{\mbld}{cmbx10 scaled 800}
\newfont{\cab}{cmsy10 scaled 1200}
\newfont{\scab}{cmsy10 scaled 1000}
\newfont{\bcall}{cmbsy10 scaled 1200}

\begin{document}
\title{A Topological Field Theory for the triple Milnor linking coefficient} 
\author{Franco Ferrari}
\email{ferrari@fermi.fiz.univ.szczecin.pl}
\affiliation{Institute of Physics and CASA*, University of Szczecin,
  ul. Wielkopolska 15, 70-451 Szczecin, Poland}
\author{Marcin R. Pi\c{a}tek}
\email{piatek@fermi.fiz.univ.szczecin.pl}
\affiliation{Institute of Physics and CASA*, University of Szczecin,
  ul. Wielkopolska 15, 70-451 Szczecin, Poland}
\affiliation{Bogoliubov Laboratory of Theoretical Physics,
Joint Institute for Nuclear Research, 141980, Dubna, Russia}
\author{Yani Zhao}
\email{yanizhao@fermi.fiz.univ.szczecin.pl}
\affiliation{Institute of Physics and CASA*, University of Szczecin,
  ul. Wielkopolska 15, 70-451 Szczecin, Poland}
\begin{abstract}
The subject of this work is a three-dimensional topological field theory with a
non-semisimple group of gauge symmetry and observables
consisting in the holonomies of connections around three closed loops.
The connections are a linear combination of gauge
potentials with coefficients containing a set of one-dimensional
scalar fields. It is checked that these observables are both metric
independent and gauge invariant. The gauge invariance is achieved by
requiring non-trivial gauge transformations in the scalar field
sector.
This topological field theory is solvable and has only a relevant
amplitude which has been computed exactly. From this amplitude it is
possible to isolate a topological invariant which is Milnor's triple
linking coefficient. The topological invariant obtained in this way is
in  the form of a sum of multiple contour integrals. The contours coincide
with the trajectories of the three loops mentioned before. The
introduction of the
one-dimensional scalar field is necessary in order to reproduce
correctly the particular path ordering of the integration over the
contours which is present in the triple Milnor linking coefficient. 
This is the first example of a local topological gauge field theory
that is solvable and can be associated to a topological invariant of
the complexity of the triple Milnor linking coefficient.
After eliminating the scalar fields, the topological
field theory of \cite{LealPineda} is recovered. This model
 is consistent whenever any pair of loops has
vanishing Gauss linking number
and the gauge potentials are invariant under abelian gauge
transformations. Some of its 
observables
 are not local.
With the addition of the one-dimensional scalar
fields a new  topological field theory has been achieved that
is consistent independently of the
way in which the loops are entangled and the observables are
local. Moreover, the gauge symmetry is much richer, because it is
based on a non-semisimple non-abelian  group.
\end{abstract}
\maketitle
\section{Introduction}\label{sec:intro}
The correspondence between three-dimensional topological field
theories and topological invariants of knots and links is well known
after the seminal paper of \cite{witten}.
From the amplitudes of Chern--Simons field theories it is for instance
possible to isolate invariants which are in the form of a sum of
multiple contour 
integrals \cite{GMM,Labastida}. Each contour integral appearing in
invariants of this kind 
may be explicitly represented as follows:
\begin{equation}
{\cal I}=\int_{a_1}^{b_1}ds_1\cdots\int_{a_n}^{b_n}ds_nf(s_1,\ldots,s_n)
\end{equation}
where $s_1,\ldots,s_n$ represent
the variables that parametrize the closed contours.
Often the integration is path ordered, which means that for some of the
pairs $s_i,s_j$ of variables it is requested that $s_i\le s_j$.
In this work the attention will be focused on invariants of this type,
which are particular cases of the
so-called  numerical knot and link invariants. The problem that will
be addressed here and that has been formulated in much more details in
Ref.~\cite{FFNOVA}, can be summarized 
by the following fundamental question:

{\it Given a particular numerical knot or link invariant expressed as a
sum of multiple contour integrals,  is it possible to find a solvable
and local topological field theory, characterized by an amplitude that
is proportional to that invariant or to a function of it?}

That posed above is not just a theoretical question. Topological field
theories 
that can be associated only to a particular invariant are one of the
main tools in studying the physics of knotted and linked polymer rings
\cite{edwards}.
Applications can also be found in other systems in which quasi
one-dimensional ring-shaped objects play a relevant role. This is the
case of magnetic lines on the surface of the Sun, which are heavily
entangled and give rise to complicated topological configurations
\cite{PriestForbes}. 
As the observations point out,  the probability of a coronal mass
ejection is growing  
with the increasing of the topological complexity \cite{Cranmer,TitovDemoulin}.
An invariant associated to a topological field theory with a
particular non-semisimple group of local symmetries derived in
\cite{FFJMP} was independently studied in connection with the solar
magnetic fields in \cite{HornigMayer}. Many other
invariants can be obtained simply by choosing  different
non-semisimple groups. An example of this strategy can be found in
Ref.~\cite{FFNOVA}, in which an invariant
describing the topological states of
a link composed by four knots has been derived.

While topological field theories with non-semisimple gauge groups have
been very successful in establishing a correspondence between
numerical link invariants and topological field theories, an important
issue is still left unsolved. Up to now, in fact, it was possible to
obtain only link invariants
containing contour integrals that are not path-ordered.
Unfortunately, all the known knot invariants and most of the  link
invariants that can be cast in the form of multiple contour integrals
require path-ordering.  

A breakthrough toward the solution of this issue is the work of
Ref.~\cite{LealPineda}, in which the case of  the triple Milnor
linking coefficient $\bar\mu(1,2,3)$ has been treated.
The only drawback of the topological field theory constructed in
\cite{LealPineda} is that some of its observables are non-local,
because they contain a
bilocal vector density. For this reason, 
the theory cannot be easily applied to polymer physics and  its full
gauge symmetry remains hidden. 
A full gauge symmetry is however required in order to deal with the
spurious degrees of freedom due to gauge invariance.
To make the model of \cite{LealPineda}
local, we start from 
a simple observation. Let us consider the path-ordered double integral
$A=\int_a^bds\int_a^sdtf(s,t)$. With the help of
a Heaviside $\theta-$function,  $A$
can be rewritten in the form
$A=\int_a^bds\int_a^bdt\theta(s-t)f(s,t)$. The crucial point is that the
$\theta-$function is the propagator of the "topological"
one-dimensional field theory
$S_\alpha=\int_{-\infty}^{+\infty}d\eta\alpha(\eta)\frac{d\alpha(\eta)}{d\eta}$. This
theory is topological in the sense that it is invariant under
reparametrization 
of the infinite line $\mathbb R$.
We show here that the topological field theory with non-local observables of
\cite{LealPineda} may be converted into a local one thanks to the
introduction of a suitable set of $\alpha-$fields. It has
been possible to prove that this local version is invariant under a
non-semisimple gauge group of symmetry like the theories discussed in
\cite{FFJMP, FFNOVA}. Using the results of these previous works, 
in particular the fact that for gauge transformations like those
considered here the Faddeev-Popov determinant is trivial,
the
partition function of the topological field theory associated to the
triple Milnor linking coefficient is explicitly computed.
\section{Conventions}
In the following Greek letters $\mu,\nu,\rho,\dots=1,2,3$ will be used
to denote the spatial indices on the flat three dimensional space
${\mathbb R}^3$.
The position of a point  $x$ on
${\mathbb R}^3$ will be given by specifying its cartesian coordinates
$x^\mu$.
Latin letters $i,j,k,\ldots=1,2,3$ will be reserved
for the indices of the internal symmetries and for labeling the three
closed trajectories $P_1,P_2,P_3$.
Throughout this paper the convention of summing over repeated indices
will be followed. In the case in which this will not be possible,
barred indices $\bar \imath,\bar \jmath,\bar k,\ldots=1,2,3$ will be adopted.
For instance, $A^{\bar \imath}B_{\bar \imath}$ is the product of the $\bar \imath-$th
components of the vectors $\boldsymbol A$ and $\boldsymbol B$, while
their vector product is defined as: $\boldsymbol A\cdot \boldsymbol
B=A^iB_i\equiv \sum\limits_{\bar \imath=1}^3A^{\bar \imath}B_{\bar \imath}$
where the symbol $\equiv$ denotes equivalence.

The trajectories $P_i$, $i=1,2,3$, will be represented by curves
$x^\mu_i(s)$ parametrized by means of their arc-lengths. It will be
assumed that all three loops have the same length $L$, so that $0\le
s\le L$.

For the formulation of the topological field theory presented here the two
triplets of vector fields
$A^i_\mu(x)$ and $a_{i\mu}$
will be needed. In addition, we introduce the set of one-dimensional
scalar field 
$\alpha_{ij}(\eta)$, where $-\infty<\eta<+\infty$.
For future purposes we define also the following three external sources:
\begin{eqnarray}
  T_i^{\mu x}&=&\oint_{P_i}dx_i^\mu\delta^{(3)}(x-x_i)=
\int_0^Lds{\dot x}_i^\mu(s)\delta^{(3)}(x-x_i(s))
\label{dist1}\\
T_i^{\{\mu x,\nu y\}}&=&\oint_{P_i}
dx_i^\mu\int_0^xdx_i^{\prime\nu}\delta^{(3)}(x-x_i)
\delta^{(3)}(y-x_i')\label{dist2}
\end{eqnarray}
and
\begin{equation}
\xi_{ij}(\eta)=\int_0^Lds\delta(\eta -s){\dot
  x}_i^\mu(s)a_{j\mu}(x_i(s))\label{dist3}
\end{equation}
Finally, the expression of the triple Milnor linking  coefficient $\bar
\mu(1,2,3)$~\cite{Milnor}, which is able to distinguish the
topological states of a 
link composed by three loops, is provided below \cite{Leal}:
\begin{equation}
\bar \mu(1,2,3)=-\frac12\int d^3x\epsilon^{\mu\nu\rho}\tilde
a_{1\mu}\tilde a_{2\nu} \tilde
a_{3\rho}+\frac12\epsilon^{ijk}\int d^3x\int d^3y T_i^{\{\mu x,\nu y\}}
\tilde a_{j\mu}(x)\tilde a_{k\nu}(y)\label{mti}
\end{equation}
where $\epsilon^{\mu\nu\rho}$ and $\epsilon^{ijk}$ are completely
antisymmetric tensors satisfying the convention $\epsilon^{123}=1$ and
\begin{equation}
\tilde a_{i\mu}(x)=\frac {\epsilon_{\mu\nu\rho}}{4\pi}\int_0^Lds{\dot
  y}_i^\rho(s) \frac{(x-y_i(s))^\nu}{|x-y_i(s)|^3}\label{fieldstilda}
\end{equation}
Using the definition of the bilocal density $T_i^{\{\mu x,\nu y\}}$ of
Eq.~(\ref{dist2}), the Milnor linking coefficient may be
explicitly expressed in terms of contour integrals over the loops $P_i$:
\begin{equation}
\bar \mu(1,2,3)=-\frac12\int d^3x\epsilon^{\mu\nu\rho}\tilde
a_{1\mu}\tilde a_{2\nu} \tilde a_{3\rho}+\frac12\sum_{\bar
  \imath=1}^3\epsilon^{\bar \imath jk}\int_0^Lds{\dot x}^\mu_{\bar
  \imath}(s)\int_0^sdt{\dot x}^\nu_{\bar
  \imath}(t)
\tilde a_{j\mu}(x_{\bar \imath}(s))\tilde a_{k\nu}(x_{\bar
  \imath}(t))\label{Milnortripleinv} 
\end{equation}
It is easy to check  that the quantity $\bar \mu(1,2,3)$ in
Eq.~(\ref{Milnortripleinv}) coincides up to an overall constant factor
with the quantity $S^1(1,2,3)$ appearing in Eq.~(16) of
Ref. \cite{LealPineda}. 
\section{The topological field theory}\label{sec2}
Let us consider the topological field theory defined by the action
\begin{eqnarray}
  S&=&\int d^3x
  \epsilon^{\mu\nu\rho}
  \left\{
4A^i_\mu\partial_\nu a_{i\rho}+\frac23\lambda\epsilon^{ijk} a_{i\mu}
a_{j\nu} a_{k\rho}
\right\}
-2\int d^3xT_i^{\mu x} A_\mu^i(x)\nonumber\\
&+&2\lambda\sum_{\bar\imath=1}^3\epsilon^{\bar \imath
  jk}\int_{-\infty}^{+\infty}d\eta\left[ 
\frac{\alpha_{\bar \imath j}{\dot \alpha}_{\bar\imath
    k}}{2}-\xi_{\bar\imath j}\alpha_{\bar\imath k}
  \right]\label{action}
\end{eqnarray}
The above action is manifestly invariant under diffeomorphisms in the
${\mathbb R}^3$ space and under reparametrizations of the $\eta$
variable.
In addition, it is possible to show that $S$ is invariant under the
set of
gauge transformations:
\begin{eqnarray}
  A_\mu^i(x)&\longrightarrow& A_\mu^i(x)+\partial_\mu\Omega^i(x) +\lambda\left(
\frac 12\omega_j(x)\partial_\mu\omega_k(x)+\omega_j(x) a_{k\mu}(x)
\right)\epsilon^{ijk}\label{gtone}\\
a_{i\mu}(x)&\longrightarrow&a_{i\mu}(x)+\partial_\mu\omega_i(x)\label{gttwo}\\
\alpha_{ij}(\eta)&\longrightarrow&\alpha_{ij}(\eta)-
\int_0^Lds\theta(\eta-s)\frac{d\omega_j(x_i(s))} {ds} \label{gtthree}
\end{eqnarray}
Here $\theta(\eta-s)$ is the Heaviside theta function and
the $\omega_i(x)$'s are arbitrary functions of the
point $x\in{\mathbb R}^3$. The functions $\Omega^i(x)$ may be
split into a single-valued and a multi-valued contribution  as follows:
\begin{equation}
\Omega^i(x)=\Omega^i_s(x)+\Omega^i_{m1}(x)+\Omega^i_{m2}(x)\label{thetavars}
\end{equation}
The $\Omega^i_s(x)'$s, $i=1,2,3$ are single-valued
 functions, while
the $\Omega^i_{m1}(x)'$s and the $\Omega^i_{m2}(x)'$s have a special
form dictated by the requirement of gauge invariance:
\begin{eqnarray}
\Omega^{\bar\imath}_{m1}(x)&=&2\lambda\epsilon^{\bar \imath
  jk}\int_{x_{\bar\imath}(0)}^x dz_{\bar
  \imath}^\mu(t)\left[
  a_{j\mu}(z_{\bar\imath}(t))
  +\frac 12\frac{\partial\omega_j(z_{\bar\imath}(t))}{\partial
    z_{\bar\imath}^\mu(t)} 
  \right]\left[
\omega_k(z_{\bar\imath}(t))-\omega_k(z_{\bar\imath}(0)) 
\right]\label{theta1}\\
\Omega^{\bar\imath}_{m2}(x)&=&-\lambda\epsilon^{\bar \imath
  jk}\int_{x_{\bar\imath}(0)}^x dz_{\bar
  \imath}^\mu(t)\left[
a_{\mu k}(z_{\bar
  \imath}(t))+\frac 12\frac{\partial\omega_k(z_{\bar\imath}(t))}{ \partial
  z_{\bar\imath}(t)} 
  \right]\omega_j(z_{\bar\imath}(t))\label{theta2}
\end{eqnarray}
In Eqs.~(\ref{theta1}) and (\ref{theta2}) $z_i^\mu(t)$ is an
arbitrary curve joining the point $x_i(0)$ belonging to the loop $P_i$
to the generic point $x\in{\mathbb R}^3$. It is easy to check that 
$\Omega^i_{m1}(x)$ and $\Omega^i_{m2}(x)$ are multi-valued
functions depending on the choice of the trajectory $z_i^\mu(t)$. We
note also  that $\Omega^i_{m2}(x)$ satisfies the 
identity:
\begin{equation}
\oint_{P_i}dx_i^\mu\left[
\partial_\mu\Omega^{\bar\imath}_{m2}+\lambda\epsilon^{\bar\imath
  jk}\left(
\frac 12\omega_j\partial_\mu\omega_k+\omega_ja_{\mu k}
\right)
\right]=0\label{impid}
\end{equation}
In words, the above relation means that the non-linear contributions
in the $\omega_i'$s appearing in the transformation (\ref{gtone}) of
the fields $A^i_\mu(x)$ is exactly canceled by the contribution due to
$\Omega_{m2}^i(x)$  when the fields $A^i_\mu$ are integrated along the
loops $P_i$.
To prove the invariance of the action (\ref{action}) under the gauge
transformations (\ref{gtone})--(\ref{gtthree}) it is convenient to
split $S$ into four contributions:
\begin{equation}
S=S_{top}+S_\alpha+S_{source1}+S_{source2}\label{actionnewdef}
\end{equation}
where $S_{top}$ is the "bulk action" on ${\mathbb R}^3$
\begin{equation}
S_{top}=\int d^3x
  \epsilon^{\mu\nu\rho}
  \left\{
4A^i_\mu\partial_\nu a_{i\rho}+\frac23\lambda\epsilon^{ijk} a_{i\mu}
a_{j\nu} a_{k\rho}
\right\}
\end{equation}
and $S_\alpha$ is the action of the one-dimensional fields
$\alpha_{ij}(\eta)$:
\begin{equation}
S_\alpha=\lambda\sum_{\bar\imath=1}^3\epsilon^{\bar\imath
  jk}\int_{-\infty}^{+\infty} d\eta \alpha_{\bar\imath
  j}{\dot\alpha}_{\bar \imath k}
\end{equation}
Finally, $S_{source1}$ and $S_{source2}$ take into account the source
terms in Eq.~(\ref{action}) and may be explicitly written as follows:
\begin{eqnarray}
S_{source1}&=&-2\sum_{\bar\imath=1}^3\int_0^Lds{\dot
  x}_{\bar\imath}^\mu(s)A_\mu^{\bar\imath }(x_{\bar\imath }(s)) \\
S_{source2}&=&-2\lambda\sum_{\bar\imath=1}^3\epsilon^{\bar\imath jk}\int_0^Lds{\dot
  x}_{\bar\imath}^\mu(s) a_{j\mu}(x_{\bar\imath}(s))\alpha_{\bar\imath
k}(s)
\end{eqnarray}
It is possible to check  that $S_{top}$ is
fully gauge invariant, i.~e.:
\begin{equation}
S_{top}\longrightarrow S_{top}\label{stopinv}
\end{equation}
To prove Eq.~(\ref{stopinv}) 
we have used
the fact that the gauge
fields $a_{i\mu}$ and $A^i_\mu$ vanish at infinity sufficiently fast
and the identity $\epsilon^{\mu\nu\rho}\int
d^3x\Omega^i\partial_\mu\partial_\nu a_{i\rho}=0$.
This identity is verified because the spatial components of $a_{i\mu}$
are not multi-valued.
Next, we consider the combination $S_\alpha+S_{source2}$ which is not fully
invariant and
 transforms as shown below:
\begin{eqnarray}
S_\alpha+S_{source2}&\longrightarrow& S_\alpha+S_{source2}+
2\lambda\sum_{\bar\imath=1}^3\epsilon^{\bar\imath jk}\int_0^Lds\Bigg[
{\dot
  x}_{\bar\imath}^\mu(s)a_{j\mu}(x_{\bar\imath}(s))\nonumber\\
&+&
\frac12\frac{d\omega_j(x_{\bar\imath}(s)) 
}{ds}\Bigg]
\left[\omega_k(x_{\bar\imath}(s))-
\omega_k(x_{\bar\imath}(0))\right]
\label{salphasst} 
\end{eqnarray}
The unwanted terms violating gauge invariance are canceled exactly by
the transformation of the term $S_{source1}$ if the functions
$\Omega^i(x)$ appearing in Eq.~(\ref{gtone})
are chosen  as in
Eqs.~(\ref{thetavars})-(\ref{theta2}). In this case, in fact, a
straightforward calculation shows that:
\begin{eqnarray}
S_{source1}&\longrightarrow& S_{source1}-
2\lambda\sum_{\bar\imath=1}^3\epsilon^{\bar\imath jk}\int_0^Lds\Bigg[
{\dot
  x}_{\bar\imath}^\mu(s)a_{j\mu}(x_{\bar\imath}(s))\nonumber\\
&+&
\frac12\frac{d\omega_j(x_{\bar\imath}(s)) 
}{ds}
\Bigg]\left[\omega_k(x_{\bar\imath}(s))-
\omega_k(x_{\bar\imath}(0))\right]\label{source1transf} 
\end{eqnarray}
Eq.~(\ref{source1transf}) has been obtained by taking into account
Eq.~(\ref{impid}). Clearly, in the action $S$ of
Eq.~(\ref{actionnewdef}) the non-invariant
contributions appearing after a gauge transformation in the right hand
side of equations (\ref{salphasst}) and (\ref{source1transf}) vanish
identically. 
As a conclusion, the action $S$ is gauge invariant.
\section{The Milnor invariant}
In the rest of this Section we will concentrate our attention
to the partition function of the theory, which is given by:
\begin{equation}
{\cal Z}=
\int{\cal D}\alpha_{ij}{\cal D}a_{i\mu}{\cal D} A^i_\mu e^{-iS}\label{pf}
\end{equation}
The above gauge field theory requires the introduction of
a gauge fixing, like for instance the Lorentz gauge. It has been shown
in \cite{FFJMP} that nonlinear gauge transformations like those of
Eq.~(\ref{gtone}) give rise to trivial Faddeev--Popov determinants, in
which the ghosts are decoupled from the gauge fields. For this reason,
the ghost sector can be ignored.

The connection with Ref.~\cite{LealPineda} is obtained after
eliminating the one-dimensional fields $\alpha_{ik}$. After performing a
Gaussian integration, it is possible to prove the
following identity:
\begin{equation}
\int {\cal
  D}\alpha_{ij}\exp\left\{-2\lambda
i\epsilon^{\bar\imath jk}\sum_{\bar\imath=1}^3
\int_{-\infty}^{+\infty}d\eta\left[
\frac{\alpha_{\bar\imath j}{\dot \alpha}_{\bar\imath
    k}}2-\xi_{\bar\imath j}\alpha_{\bar \imath k}
    \right] \right\}=e^{2i\lambda I}
\end{equation}
where
\begin{equation}
  I=\sum_{\bar\imath=1}^3\epsilon^{\bar \imath jk}\int_0^Lds\int_0^Ldt
  \theta(s-t) {\dot x}_{\bar\imath}^\mu(s)
        {\dot x}_{\bar\imath}^\nu(t) a_{j\mu}(x_{\bar\imath}(s))
        a_{k\nu}(x_{\bar\imath}(t) )
\end{equation}
and the Heaviside function $\theta(s-t)$ is nothing but the propagator
of the fields $\alpha_{ij}$.
The quantity $I$ can be cast in the form of a double volume integral
in ${\mathbb R}^3$:
\begin{equation}
I=\epsilon^{ijk}
\int d^3x\int d^3y\epsilon^{ijk} T_i^{\{\mu x,\nu y\}}a_{j\mu}(x)a_{k\nu}(y)
\end{equation}
Thus,
the partition
function in Eq.~(\ref{pf}) may be rewritten as follows:
\begin{equation}
{\cal Z}=\int{\cal D}A_\mu^i{\cal D}a_{i\mu} e^{-iS'}\label{pfprime}
\end{equation}
where
\begin{eqnarray}
  S'&=&\int d^3x\epsilon^{\mu\nu\rho}\left\{
4 A_\mu^i\partial_\nu a_{i\rho}+\frac 23\lambda\epsilon^{ijk}a_{i\mu}a_{j\nu}a_{k\rho}
\right\}\nonumber\\
&-&2\int d^3x T_i^{\mu x}A_\mu^i(x)+2\lambda\int d^3x\int d^3y T_i^{\{
  \mu x,\nu y\} } a_{j\mu}(x)a_{k\nu}(y)
\end{eqnarray}
In this way the topological field theory discussed in
\cite{LealPineda} has been recovered. 
In the partition function (\ref{pfprime}) the gauge fields $A_\mu^i$ are
playing the role of Lagrange multipliers imposing the condition:
\begin{equation}
\epsilon^{\mu\nu\rho}\partial_\nu a_{i\rho}(x)=\frac 12 T_i^{\mu x}\label{staeq}
\end{equation}
These fields can be integrated out giving as a result:
\begin{equation}
{\cal Z}=\int {\cal D}a_{i\mu}(x)e^{-iS^{\prime\prime}}
\delta\left(
4\epsilon^{\mu\nu\rho}\partial_\nu a_{i\rho}(x)-2T_i^{\mu x}
\right)
\end{equation}
where
\begin{equation}
S^{\prime\prime}=\int d^3x\frac
23\lambda\epsilon^{\mu\nu\rho}\epsilon^{ijk} a_{i\mu}
a_{j\nu}a_{k\rho}
+2\lambda\int d^3x\int d^3y \epsilon^{ijk}T_i^{\{\mu x,\nu
  y\}}a_{j\mu}(x)a_{k\nu}(y)
\end{equation}
After a further integration over the fields $a_{i\mu}$, the partition
function becomes ${\cal Z}=e^{iS^{\prime\prime}}$, where
$S^{\prime\prime}$ is computed at the solutions of
Eq.~(\ref{staeq}). These solutions 
are nothing but the
$\tilde a_{i\mu}'$s 
provided in
Eq.~(\ref{fieldstilda}) apart from a proportionality factor $-2$.
Comparing the form of $S^{\prime\prime}$ with that of the Milnor
linking coefficient $\bar\mu(1,2,3)$ of Eq.~(\ref{mti}), it is clear
that they coincide and it is possible to conclude that:
\begin{equation}
{\cal Z}=e^{2i\lambda\bar\mu(1,2,3)}
\end{equation}
\section{Conclusions}
The topological field theory defined by the action (\ref{action}) is
invariant under diffeomorphisms on ${\mathbb R}^3$ and
reparametrizations of the variable $\eta\in\mathbb R$. As it has been
shown in Section~\ref{sec2}, it is also gauge invariant. Finally, it
is exactly solvable. Its partition function consists essentially in
the Milnor linking coefficient $\bar\mu(1,2,3)$. 
This is the first
time that a local topological field theory has been constructed whose 
partition function can be computed in closed form and is associated to a
topological invariant of the complexity of the Milnor linking
coefficient.
Applications of this topological field theory to polymer physics are
currently work in progress. 
\section{Acknowledgments}
The support of the Polish National Center of Science,
scientific project No. N~N202~326240, is gratefully acknowledged.

\end{document}